\documentclass[ams, prb, twocolumn, amssymb, floatfix, longbibliography]{revtex4-1}
\usepackage{amsmath}
\usepackage{tabularx}
\usepackage{bm}
\usepackage{euscript}
\usepackage{graphicx}
\usepackage{color}
\usepackage{amsfonts}
\usepackage{exscale}
\usepackage{amsbsy}
\usepackage{subfigure}
\usepackage{textcomp}
\usepackage{comment}
\usepackage{slashed}
\usepackage{mathrsfs}
\usepackage{hyperref}
\usepackage{bbold}

\newcommand{\nn}{\nonumber}

\newcommand{\beq}{\begin{equation}}
\newcommand{\eeq}{\end{equation}}
\newcommand{\be}{\begin{eqnarray}}
\newcommand{\ee}{\end{eqnarray}}

\begin{document}

\title{Self-duality of the integer quantum Hall to insulator transition: composite fermion description}
\author{Prashant Kumar$^{1}$, Yong Baek Kim$^{2}$, S. Raghu$^{1}$}
\affiliation{$^{1}$Stanford Institute for Theoretical Physics, Stanford University, Stanford, California 94305, USA}
\affiliation{$^2$Department of Physics and Centre for Quantum Materials, University of Toronto, Toronto, Ontario M5S 1A7, Canada}
\date{\today}

\begin{abstract}
The integer quantum Hall to insulator transition (IQHIT) is a paradigmatic quantum critical point.  Key aspects of this transition, however, remain mysterious, due to the simultaneous effects  of quenched disorder and strong interactions.  We study this transition using a composite fermion (CF) representation, which incorporates some of the effects of interactions.  As we describe, the transition also marks a IQHIT of CFs: this suggests that  the transition may exhibit  `self-duality'.  
We show the explicit equivalence of the electron and CF Lagrangians at the critical point via the corresponding 
non-linear sigma models, revealing the self-dual nature of the transition.  
We show analytically  that the resistivity tensor at the critical point is $\rho^c_{xx} = \rho^c_{xy} =  \frac{h}{e^2}$, which are consistent with the expectations of  self-duality, and in rough agreement with experiments.  
\end{abstract}

\maketitle

\section{Introduction}
In two dimensions, the dc electrical resistivity  can be {\it universal} at a continuous transition between two distinct electronic phases of matter.  An example occurs at the transition from an integer quantum Hall state at filling fraction $\nu = 1$, to insulator in a disordered two dimensional electron gas in a strong perpendicular magnetic field.  Experimental studies have confirmed that the resistivity tensor at this critical point is $\rho^c_{xx} \approx \rho^c_{xy} \approx  \frac{h}{e^2}$.\cite{Shahar1995,Shahar1997,Yang2000,Dunford2000}   While both the integer quantum Hall and insulating phases can be adiabatically deformed to free fermions, the critical point separating them necessarily involves strong interactions.
It therefore remains a fundamental and open challenge to construct a quantum theory of electrical transport at the  transition.  

Often, however, strongly interacting systems can have alternate `dual' descriptions in terms of degrees of freedom that provide a nearly free quasiparticle framework.  In the quantum Hall context, interacting electrons in a partially filled  Landau level can be mapped, via   flux attachment, to particles carrying charge and flux known as composite fermions (CFs).\cite{Jainbook, Lopez91, Kalmeyer1992, Halperin1993}  While CFs are coupled to a dynamical gauge field, they are described in terms of a filled Fermi sea, which acts to mitigate much of the strong gauge fluctuation effects and to allow for physically motivated,  mean field Fermi liquid like descriptions.  The latter, with the inclusion of quenched disorder effects, provide us with an alternate view of the quantum Hall to insulator transition, one which, as we shall see, is amenable to explicit calculation of transport quantities.  

In this paper, we study the integer quantum Hall to insulator transition (IQHIT) from such a CF viewpoint.  As we explain, a careful treatment of quenched disorder  effects near the IQHIT,  in the CF language, leads to the conclusion that the CFs themselves sit at an integer quantum Hall transition.  Since the transition corresponds to a IQHIT in both electron and CF representations, it has two identical manifestations involving very different degrees of freedom, suggestive of an underlying self-duality.


At a self-dual critical point, the electron and composite fermion Lagrangians must take the same form.  Previous studies of the IQHIT transition in electron coordinates led to the conclusion that the effective theory is a non-linear sigma model (NLSM) with a non-zero theta term\cite{Levine1983,Khmelnitskii1983, Pruisken1984}.  Further evidence suggesting such a theory came from the analysis of network models of quantum Hall transitions, which can in turn be mapped onto spin-Pierels transitions of coset models; the theta term required for the spin-Pierels transition is identified with the theta term associated with the IQHIT.\cite{Affleck1986,Lee1994}

In this paper, we explicitly construct the effective theory for the IQHIT in the  CF representation.  We show that up to an important sign, it has the same form as the electron Lagrangian.  Thus, we establish the self-dual nature of the transition, as seen from electron and CF theories.  We provide further evidence of self-duality by considering CF network models for the transition, and show that they are in the same universality class as the electron network model for the IQHIT.  
We then conclude from this self-duality, that at the transition, the critical dc resistivity tensor is $\rho_{xx}^{c} = \rho_{xy}^c = h/e^2$, which is nearly consistent with experimental observations.  
In previous approaches to this problem, which studied the transition in electron coordinates, it was argued that $\sigma_{xy}^c = e^2/2h$.  However, the longitudinal conductivity was left unspecified.  Our key result here is that by studying the transition in CF as well as electron coordinates, both components of the conductivity tensor are uniquely determined at criticality.  


%
%

This  paper is organized as follows. In section \ref{sec:cf_with_disorder}, we present the IQHIT from the perspective of CF mean-field theory. Section \ref{sec:NLSM} describes the CF non-linear sigma model.  
In section \ref{sec:network_model}, we construct CF network models and contrast them with the original network model of the IQHIT described in terms of electrons.  
We discuss the implications of self-duality in section \ref{sec:self_duality}. 
In appendix \ref{topological_term_derivation}, we present a self-contained derivation of the non-linear sigma model, including the topological term. 
In appendix \ref{ff_duality}, we discuss the self-duality from the complementary (and as we show, equivalent) perspective of Dirac composite fermion theory.\cite{Son2015}

\section{Motivation: Composite fermions with quenched disorder \label{sec:cf_with_disorder}}
In the vicinity of the IQHIT, the low energy electron degrees of freedom are governed by 

\begin{gather}
Z[A] = \int D \bar \psi D \psi\ e^{i \int \mathscr{L}[A]},\\ 
 \mathscr{L}[A]= \bar\psi\left[\hat K_{A} +\mu + V(\bm r)\right]\psi+\cdots
\end{gather}
where $\hat K_{A} = i D^A_t + \frac{1}{2m} D^A_jD^A_j$, $D_\mu^A = \partial_\mu-iA_\mu$, $A_{\mu}$ is an external vector potential with $\nabla\times \bm A = B$, with $B$ a uniform external magnetic field. The chemical potential $\mu$ sets the density of electrons to be at filling fraction $\nu=1/2$, and $V(\bm r)$ is a quenched random potential taken from a probability distribution with  variance 
\begin{equation}
\overline{V(\bm r) V(\bm r')} = \Delta f(\vert \bm r - \bm r' \vert),
\end{equation}
where $f(r)$ is some smooth function of position.  In GaAs, modulation doping introduces impurities in a different layer than that of the 2d electron gas.  As a consequence, the disorder potential varies slowly compared to the magnetic length.  Below, it will be convenient to allow for a nonzero uniform component of $V(\bm r)$, which simply corresponds to a uniform chemical potential shift away from $\nu = 1/2$.  

To formulate the problem in composite fermion coordinates,  we introduce composite fermion fields $\bar f, f$, a dynamical $U(1)$ gauge field $a_{\mu}$ and express the {\it same} $Z[A]$ above in terms of these fields as\cite{Halperin1993}:
\begin{gather}
Z[A] = \int D \bar f D f D a\ e^{i \int \mathscr{L}_{cf}[A]} \\
\mathscr{L}_{cf}[A] = \bar f \left[\hat K_{a + A} + \mu + V(\bm r) \right] f + \frac{1}{8 \pi} \epsilon_{\mu \nu \lambda} a_{\mu} \partial_{\nu} a_{\lambda} + \cdots
\end{gather}
In the simplest mean-field approach, the composite fermion fields have the same mass as the electrons, their covariant derivatives involve the sum of the background field $A$ and dynamical field $a$, and the last term above is a Chern-Simons term that acts to attach two units of flux.  In what follows, we will employ the following shorthand
\begin{equation}
\epsilon_{\mu \nu \lambda} a_{\mu} \partial_{\nu} b_{\lambda} \rightarrow adb .
\end{equation}
Note in particular that the composite fermions encounter the same potential $V(r)$ as do the electrons. 
For later convenience, let's shift $a\rightarrow a- A$:
\begin{gather}
\label{HLR}
\mathscr{L}_{\rm cf}[A] = \bar f \left[\hat K_{a} + \mu + V(\bm r)  \right] f + \frac{1}{8 \pi} (a-A)d(a-A) + \cdots
\end{gather}

Since the disorder is long-wavelength in character, we can use linear response theory to obtain the density variation in terms of a long-wavelength compressibility:
\begin{gather}
\bar f f(\bm r)  = \bar n + \delta n(\bm r) = \bar n + \chi V(\bm r),
\label{compressibility}
\end{gather}
where $\bar n$ is the average density set by the chemical potential $\mu$, and $\chi$ is the compressibility. For a 2-dimensional electron gas, $\chi = \frac{m}{2\pi}$.  As shown in Ref. \onlinecite{KimFurusakiWenLee1994}, the long-wavelength limit of $\chi$ does not suffer from mass renormalization due to gauge fluctuations, and is expressible in terms of the {\it bare} mass.  This is certainly true of the mean-field treatment employed in the present analysis.  
Furthermore, the equation of motion of $a_0$ leads to the flux attachment constraint:
\begin{gather}
b(\bm r) = \nabla \times \bm a(\bm r) =  B-4\pi \bar ff(\bm r) 
\label{eom}
\end{gather}
%
%
%
Using Eqs. \eqref{compressibility} and \eqref{eom},  the disorder potential can be related to the effective magnetic field:
\begin{gather}
V(\bm r)  = -\frac{b(\bm r)}{2m} \label{slave_disorder_relation}.
\end{gather}
As we noted before, $V(\bm r)$ can have a uniform component, which corresponds to a deviation from $\nu=1/2$.  The above equation shows that such a deviation corresponds to a uniform component to the magnetic field $b(\bm r)$ felt by composite fermions.  
The mean-field Hamiltonian obtained from this constraint is:
\begin{align}
\mathcal{H}_{\rm cf} &= \frac{(p-a)^2}{2m} - V(\bm r)\nn\\
&= \frac{(p-a)^2}{2m} + \frac{g}{2}\frac{b(\bm r)}{2m} \label{HLR_Hamiltonian}
\end{align}
where $g=2$. This Hamiltonian describes a free particle in a random magnetic field and random potential.  Importantly, both disorder fields are obtained from a {\it single} quenched random field, namely $a_{\mu}(r)$;  the randm magnetic field and electric potential are not uncorrelated. This slaving between the two disorders occurs due to flux attachment, and  $g=2$ follows from both flux attachment, and the use of the {\it bare} mass in the long wavelength compressibility.  

Notice that the discussion above neglects the Coulomb interaction between electrons in the original problem.
When the Coulomb interaction between electrons is explicitly taken into account, the mass of the composite fermion would be 
set by such interaction energy scale near $\nu=1/2$.
On the other hand, as long as the compressibility of the composite fermion is set by the same mass, as suggested by
earlier studies of gauge field fluctuations, Eq. \eqref{HLR_Hamiltonian} with proper replacement of the mass would remain valid. 
In the subsequent analyses, we use the bare mass for simplicity.



The inclusion of slaved disorder in Eq. \eqref{HLR_Hamiltonian} has singular consequences:\cite{Wang2017,KumarTrivialPaper,Kumarsusy}
an infinitesimal amount of slaved disorder has a non-perturbative effect in inducing an order one Hall conductivity.
\begin{figure}
	\includegraphics[width=2.5in]{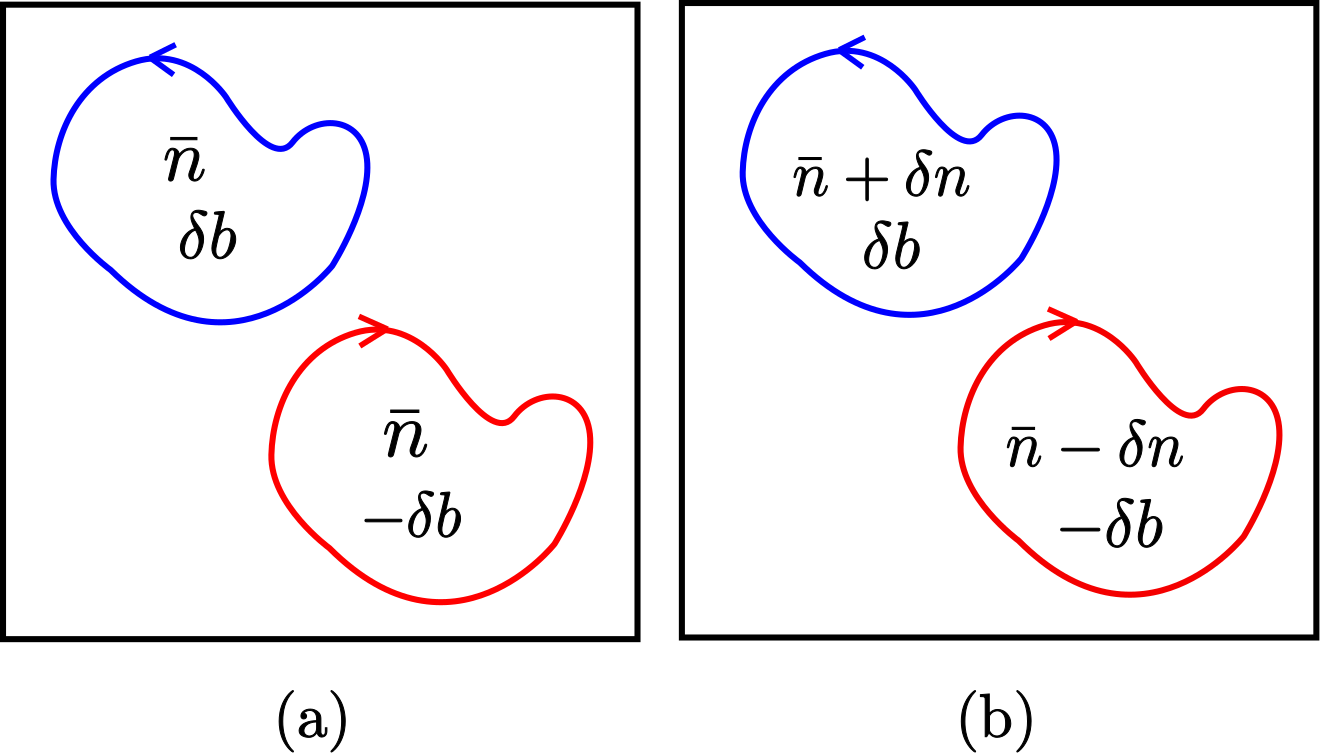}
	\caption{(Taken from Ref. \onlinecite{KumarTrivialPaper}) (a) A cartoon of a typical spatial configuration with pure flux disorder.  The system has an average density $\bar n$.  For every region with magnetic flux $\delta b$, there is an equal region with the opposite flux $- \delta b$, and hence the Hall conductivity in this system vanishes.  (b) By contrast, when fermion density and flux disorder are correlated, as in Eq. \eqref{eom}, an analogous cartoon shows that the Hall conductivity need not vanish. This is so because the filling fractions of the two regions do not completely cancel each other. As shown in Eq. \eqref{Lord_Kelvin}, this imbalance can induce to an order one Hall conductivity.  
	}
	\label{fig:blob}
\end{figure}
To see this, consider the effect of spatial inhomogeneity in 
Fig. \ref{fig:blob}.  The regions where composite fermions move counter-clockwise do not completely cancel out the regions where they move clockwise. 
As a heuristic estimate, we can average over the filling fractions of these two regions and obtain the Hall conductivity:
\begin{gather}
\sigma_{xy}^{\rm cf} \simeq -\frac{1}{2\pi} \nu^{\rm cf}_{\rm eff} \approx \frac{\delta n}{\delta b} = -\frac{1}{4\pi} \label{Lord_Kelvin}
\end{gather}
%
Note that the zero disorder limit ($\delta n, \delta b \rightarrow 0$) is singular in the CF representation: we resolve the singularity using the flux attachment constraint to find  an order one   $\sigma_{xy}^{\rm cf}$.  As shown in Ref. \onlinecite{Kumarsusy}, this heuristic estimate for $\sigma_{xy}^{\rm cf} $ is {\it exact} in the long wavelength limit.  

The fact that $\sigma_{xy}^{\rm cf}$  is a half-integer multiple of  $\frac{1}{2\pi}$ takes a special significance: states at the Fermi level must be {\it extended}, as follows from a  corollary of Laughlin's gauge argument.\cite{Laughlin1981}  
Extended states at the Fermi level, along with 2-dimensionality, and  the fact that interactions are neglected in CF mean field theory, indicate that the fermions must be at a critical point between two integer Quantum Hall states, in this case  $\nu^{\rm cf} = -1$ and $\nu^{\rm cf} = 0$.  
%
%

\begin{figure}
	\centering
	\includegraphics[width=2.3in]{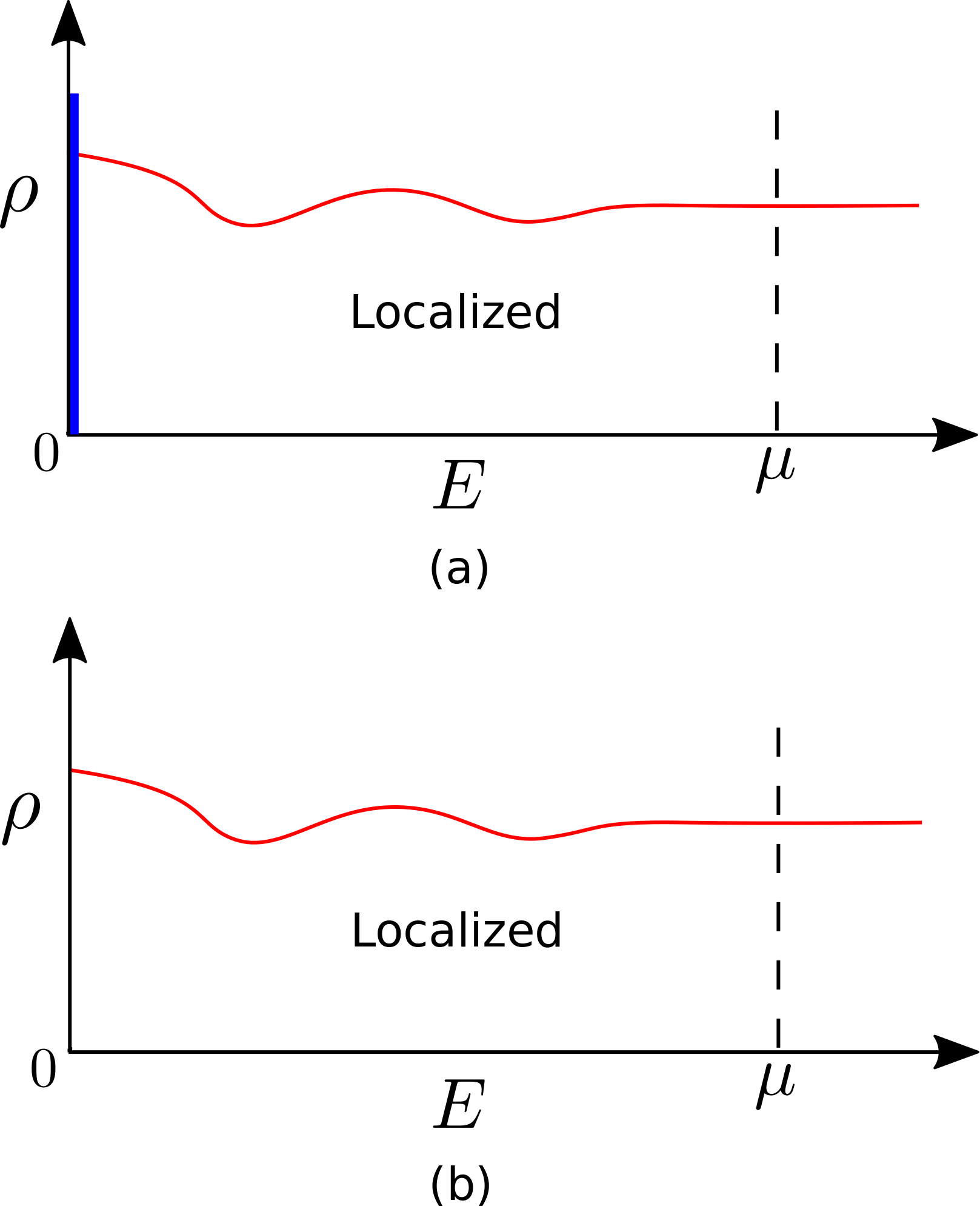}
	\caption{(Taken from Ref. \onlinecite{Kumarsusy}) A schematic for the density of states of composite-fermion Hamiltonian $\mathcal{H}_{\rm cf}$ in Eq. \eqref{HLR_Hamiltonian} for (a) $b_0 < 0$ and (b) $b_0 > 0$. Zero-energy states are present only for $b_0 < 0$ (represented by the Dirac delta-function) and contribute a Hall conductance $\sigma_{xy}^{\rm cf} = -\frac{e^2}{h}$. All positive energy extended states are assumed to have levitated up for $|b_0| \rightarrow 0$. Therefore, $b_0 = 0$ corresponds to a critical point between two integer quantum Hall states with $\sigma_{xy}^{\rm cf} = -\frac{e^2}{h}$ and $\sigma_{xy}^{\rm cf} = 0$.}
	\label{fig:IQHT_HLR}
\end{figure}

To see this more directly,  note that $\mathcal H_{\rm cf}$ in Eq. \eqref{HLR_Hamiltonian} with $g=2$, can have {\it exact} zero-energy modes even for arbitrary disorder.\cite{Aharonov1979}   The zero-energy modes will play a crucial role in the remainder of the paper.  For now, we show how they ensure that the CFs sit at an IQHIT.    A typical density of states of $\mathcal H_{\rm cf}$  is plotted in Fig. \ref{fig:IQHT_HLR}.  Without loss of generality, consider the case where all finite energy states are localized due to disorder.  
The  uniform component of $b(\bm r)$, denoted $b_0$, which detunes the system from $\nu=1/2$, will act as the tuning parameter for the IQHIT:
\begin{equation}
b(\bm r) = b_0 + \tilde b(\bm r)\label{b_separate}
\end{equation}
where, 
$\langle\tilde{b}(\bm r)\rangle = 0$. 
When $b_0  < 0$, there are  exact zero modes, the number of which is equal to the total number of flux quanta passing through the system.  The zero modes therefore act as a Landau level:  for {\it any} non-zero Fermi level, they  provide an integer Hall conductance: $\sigma_{xy}^{\rm cf} = - 1/2\pi$, where the minus sign occurs due to the fact that $b_0 < 0$. However, when $b_0 > 0$, there are no zero energy states: the system therefore has zero Hall conductance for $b_0 > 0$. Therefore, we see that 
\begin{align*}
\sigma_{xy}^{\rm cf} = \left\{ \begin{matrix}
-\frac{1}{2\pi}, & b_0 < 0\\
0, & b_0 > 0
\end{matrix}\right.
\end{align*}
and the IQHIT maps on to an IQHIT $\nu^{\rm cf}=-1\rightarrow 0$ of CFs as $b_0$ is tuned through zero.    At the critical point itself, when $b_0 = 0$, the Hall conductance is $\sigma_{xy}^{\rm cf} = -1/4\pi$ if all odd moments of disorder vanish. This conclusion holds for arbitrary disorder strength provided that the condition $g=2$ is satisfied.\cite{Kumarsusy}

To summarize this section, the $\nu=1 \rightarrow 0$ IQHIT of electrons is encapsulated as a $\nu^{\rm cf} = -1 \rightarrow 0$ IQHIT of CFs.  The only difference between the two representations arises in the sign of the Hall conductance at criticality.  This strongly suggests that the IQHIT exhibits self-duality.  
In the following two sections, we establish the self-dual nature of the transition. 

\section{Composite fermion NLSM and Self-duality \label{sec:NLSM}}
Armed with the intuition in the previous section, we now construct the low energy effective theory that governs the behavior of diffusive modes at length scales large compared to the mean-free path. 
We start with the following mean-field lagrangian of disordered composite fermions:
\begin{align}
\mathcal{L} = \psi^\dagger\left(i D_t + \mu - \frac{g}{2}\frac{b(\bm r)}{2m} + \frac{1}{2m}D_j^2\right)\psi
\end{align}
where $D_\mu = \partial_\mu - ia_\mu$ and $g=2$.

For simplicity, we take the vector potential to have a spatially uncorrelated Gaussian random distribution: 
\begin{align}
P\left[a_j(\bm r)\right] \propto \exp\left[-\frac{1}{2\Delta}\int d^2r\ a_j^2(\bm r)\right]
\end{align}
Since the $g=2$  theory is the ``square" of a Dirac lagrangian,  we can map the composite-fermion lagrangian at $g=2$ to a Dirac fermion through a fermionic Hubbard-Stratonovic transformation:
\begin{align}
\mathcal{L} &= i\psi^\dagger D_t \psi + \mu \psi^\dagger\psi + \mu\chi^\dagger\chi\nn\\
&\ \ \ \ \ \ \  + iv\psi^\dagger(D_x-iD_y)\chi + iv\chi^\dagger(D_x-iD_y)\psi
\end{align}
where $v=\sqrt{\frac{\mu}{2m}}$.

Defining $\Psi = \begin{pmatrix}\psi \\ \chi \end{pmatrix}$ and $\bar\Psi = \Psi^\dagger\sigma_z$, we can rewrite the lagrangian as that of a Dirac fermion with a random vector potential.
\begin{align}
\mathcal{L} &= i\psi^\dagger D_t \psi + iv\bar\Psi\gamma^j D_j \Psi + \mu \bar\Psi\gamma^0\Psi\label{geq2_dirac} + \mathcal O(g-2)
\end{align}
where $\gamma^0 = \sigma_z$, $\gamma^1 = i\sigma_y$ and $\gamma^2 = -i\sigma_x$, and the $\mathcal O(g-2)$ terms arise from $g-2$ corrections.  We shall estimate the leading order effect of $g-2$ terms below. 
 
Note that $\chi$ fields don't have a time derivative.  However, within composite fermion mean-field theory, this lack of time-derivative for $\chi$ has no consequence: when interactions are ignored, only spatial variations need to be considered, and frequency acts as a 	``parameter" of the theory\cite{Efetov1980, Nayaklectures}.  The disorder problem can be analyzed one frequency at a time, and for the dc limit of interest, the system for all intensive purposes behaves as a 2-component Dirac fermion in $d=2+0$ dimensions,  in a random vector potential.  

To construct the effective field theory, we follow the standard procedure described in Refs \onlinecite{Efetov1980, Nayaklectures}.  The idea is to consider disordered averaged products of retarded and advanced Green fuctions, which are immune to dephasing and capture the low energy diffusive physics.  We do so for a Dirac fermion in a random vector potential, and perform disorder averaging using the replica trick.  With $n$ copies of retarded and advanced fermions (with $n \rightarrow 0$), the theory naively has a $U(2n)$ symmetry.  However, due to a finite density of states, the imaginary parts of the retarded and advanced functions differ in sign, and the $U(2n)$ theory is actually broken down to $U(n) \times U(n)$.  The diffusive modes correspond to the Goldstone modes living in the Grassmannian manifold $U(2n)/U(n) \times U(n)$.\footnote{The $U(2n)/U(n) \times U(n)$ manifold is equivalent to the unitary class of random matrix theory, which corresponds to the case where both spin-rotation and time-reversal symmetry are broken, and the fermion density is the only conserved mode that exhibits diffusion.}   They are related to the longitudinal conductivity.  However, as we explicitly derive in appendix \ref{topological_term_derivation}, there is in addition a topological term present, which arises from the Chiral anomaly of the Dirac fermion.  This term reflects the fact that the CF theory at $g=2$ has a Hall conductivity of $-1/4\pi$, as described in the previous section.  For a detailed derivation of the theory, we invite the reader to study the appendix.  For the reader interested in the summary, we state here the resulting theory:


%
\begin{gather}
Z = \int \prod DQ^\dagger DQ\ e^{-\int d^2r\ \mathcal{L}_{\rm eff}}\nn\\
\mathcal L_{\rm eff} = \frac{\pi\sigma_{xx}^{\rm cf}}{4}\mathrm{Tr}\left[\partial Q\right]^2 + \frac{\pi \sigma_{xy}^{\rm cf}}{4} \epsilon_{ij}\mathrm{Tr}\left[Q\partial_iQ \partial_jQ\right]\label{NLSM}
\end{gather}
where $\sigma_{xx}^{\rm cf} = \frac{1}{4\pi \Delta}$ and $\sigma_{xy}^{\rm cf} = -\frac{1}{4\pi}$. $Q \equiv Q_{\alpha\beta;\sigma \sigma'}$, that represents the long wavelength diffusive modes, is a hermitian matrix with replica indices $\alpha,\beta$ and retarded/advanced indices $\sigma, \sigma'$. Also, $Q = u^\dagger \Lambda u$, where
\begin{gather*}
\Lambda \equiv \Lambda_{\alpha\beta;\sigma \sigma'} = \delta_{\alpha\beta}\sigma^z_{\sigma \sigma'},
\end{gather*}
$\sigma^z$ is the z-component of Pauli matrix and $u$ is a $2n \times 2n$ unitary operator in the replica and retarded/advanced space. 
%
%


The NLSM for composite fermions should be contrasted with the effective description of electrons undergoing an IQHIT\cite{Pruisken1984}:
\begin{equation}
\mathcal L_{\rm Pruisken} = \frac{\pi\sigma_{xx}}{4} {\rm Tr} \left( \partial Q \right)^2 + \frac{\pi\sigma_{xy}}{4} \epsilon_{ij} {\rm Tr} \left[ Q \partial_i Q \partial_j Q \right]
\end{equation}
where $\sigma_{xy} = \frac{1}{4\pi}$. The electron NLSM is identical to the composite fermion NLSM with one important difference: the two theories have relative  opposite signs for the topological term, proportional to their Hall conductivities at the critical point. 

The equivalence of electron and CF effective field theories shows that the IQHIT is a self-dual critical point.
Thus, the universal quantities such as the dc conductivity tensor and the critical exponents must also be equivalent, up to signs of the Hall conductivity, in the two theories. 

An important question to ask is whether the theory presented in this section is stable under perturbations that can arise for example from short-range or stronger disorder. To analyze this, let's consider a small deviation from $g=2$.  The most relevant term at the $g=2$  free $d=2+0$ dimensional Dirac fermion fixed point is a combination of random chemical potential and mass for the Dirac fermions:
\begin{align}
\Delta\mathcal{L} = -\frac{g-2}{4m}\ b(\bm r)\left(\bar\Psi \gamma^0 \Psi + \bar\Psi  \Psi\right)
\end{align}
A naive dimensional analysis, holding the kinetic term of  a $d=2+0$ dimensional free fermion fixed, resulting in $[\psi]=1/2$ suggests that this term is irrelevant in the RG sense: $\left[ g-2 \right] = -1$. However, the naive power counting estimate is valid only for weak disorder; $g-2$ corrections could be dangerously irrelevant as they lead to a shift in the Hall conductivity.

\section{ Self-duality and composite fermion network models\label{sec:network_model}}
In this section, we describe an alternate and more vivid manifestation of the self-duality in terms of network models.
Network models\cite{chalkercoddington} have played a  pivotal role in the study of integer quantum Hall transitions. 
They have been applied to case of a random magnetic field  in Ref. \onlinecite{PhysRevB.50.5272}. Generalizing their approach, we construct a network model for composite fermions where the additional zero mode plays a crucial role. 
This will provide a complementary insight into the self-dual nature of the IQHIT, corroborating the analysis of the previous section.  


\begin{figure}
	\centering
	\includegraphics[width=2.5in]{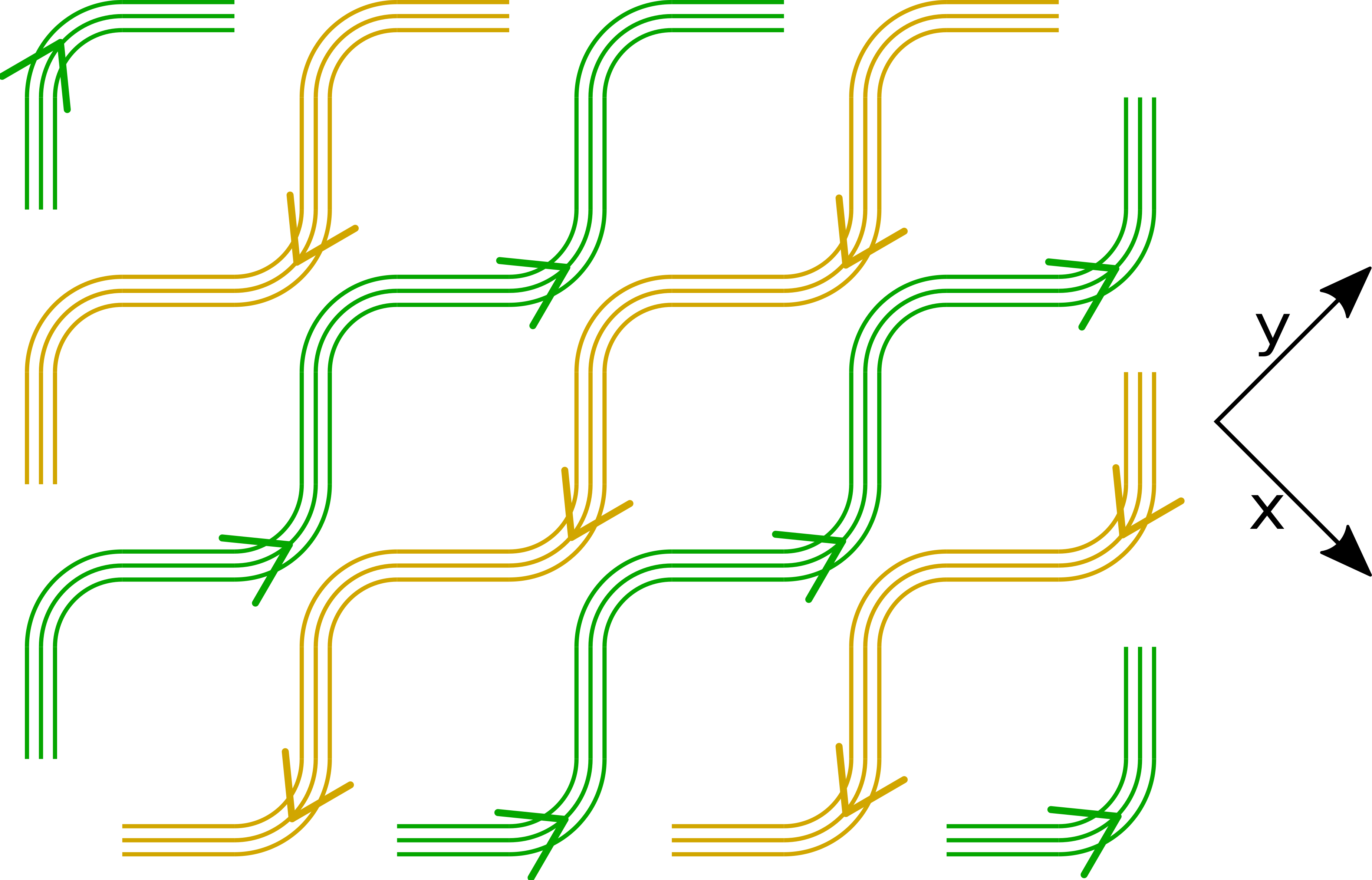}
	\caption {The network model for three chiral edge modes. The counter-propagating channels are shown using green and yellow colors. Each channel hosts $N_{\rm em}$ number of co-propagating edge modes and is mapped to a single site of the 1-dimensional spin-chain of Eq. \eqref{spin_Hamiltonian}.\cite{Affleck1986,Lee1994,PhysRevB.52.16646} On the other hand, the edge modes are mapped to the $SU(2n)$ spins at each site (with $n\rightarrow 0$ denoting the replica limit). In the figure, the $x$-direction labels the sites while the $y$-direction acts as the imaginary time axis for the spins. The disordered-averaged tunneling amplitudes between the edge modes at the nodes of the network model (the vertices of the plaquettes) translate to the inter- and intra-site couplings of the spin-chain.
	}
	\label{fig:spin_map}
\end{figure}

For fermions with a smooth random magnetic field, the $b(\bm r)=0$ contour percolates throughout the system. 
The edge states on these contours form the channels of the network model, where each channel hosts a number of
co-propagating edge modes. 
We can construct the network model by counting the number of edge modes in each channel.
To this end, let's consider two adjacent regions 1 \& 2 such that in region 1, $b = |b_0|$ and in region 2, $b = -|b_0|$. The Landau level energies for CFs at $g=2$ are given by:
\begin{align}
E_n = \left\{\begin{matrix}
(n+1)\hbar\omega_0, & \text{Region 1}\\
n\hbar\omega_0, & \text{Region 2}
\end{matrix}\right.
\end{align}
where $\omega_0 = \frac{e|b_0|}{m}$.

Thus, there is a filled Landau level at zero energy present in only the region 2. This is precisely the extra zero mode discussed in section \ref{sec:cf_with_disorder}. Additionally, the energy of $p^{th}$ Landau level in region 1 matches with the energy of $(p+1)^{th}$ Landau level in region 2. 
Each Landau level from either region contributes one edge state at the $b = 0$ boundary.\cite{PhysRevB.52.16646} 
Therefore, if the Fermi energy lies in between two bulk Landau levels,
the number of edge states in each channel in the network model will generically be an {\it odd }number $N_{\rm em}=2M+1$.

An obvious special case of such a class of network models is when $N_{\rm em} = 1$. This is identical to the original Chalker-Coddington model of electrons in the lowest Landau level. Thus, at least for the special case of $N_{\rm em} = 1$, the IQHIT of CFs and electrons are both in the same universality class.

To study the general case of the network model with odd number of edge states in each channel, 
we utilize the mapping of the network model to coupled
1-dimensional spin-chains.\cite{PhysRevB.52.16646} 
As shown in Fig. \ref{fig:spin_map}, the transport in the network model is described by random tunneling between the edge states, where $N_{\rm em}$ number of chiral
edge modes are moving in the same direction in the given channel while the direction of chiral edge states is
alternating between nearby channels. Representing the motions of alternating chiral edge modes 
as the 1+1 dimensional world lines of fermions and after taking the average over the random tunneling amplitude via the replica trick,
the resulting interacting fermion model is mapped to the 1-dimensional spin chain model, where $N_{\rm em} = 2M+1$ number of $SU(2n)$ 
spins reside at each site $x$ with $n\rightarrow 0$, denoting the replica limit. 
Here the site $x$ represents the ``space-like" positions of the chiral edge modes in the world line representation
mentioned above.
The Hamiltonian of the spin-chain is:
\begin{align}
\mathcal H_{\rm spins} &= -\sum_{x,j,k}\frac{J_{jk}}{2}\ \mathrm{Tr}\left[S_j(x)S_k(x)\right]\nn\\
&\ \ \ \ \ \ \ \ \  + \sum_{x,j} J_j\ \mathrm{Tr}\left[S_j(x+1)S_j(x)\right]\nn\\
&\ \ \ \ \ \ \ \ \ -\sum_x (-1)^x \eta \mathrm{Tr}\left[\Lambda S_j(x)\right]\label{spin_Hamiltonian}
\end{align}
where $S_j(x)$ denotes the $j^{th}$ spin on site $x$ and $j = 1, ..., N_{\rm em}$. 
$\eta \rightarrow 0^+$ and $\Lambda$ is a $2n\times 2n$ diagonal matrix: $\Lambda = \text{Diag}\left[1,1,\ldots, 1, -1,-1,\ldots,-1\right]$.
Within each site $x$, the spins are ferromagnetically coupled via $J_{jk}$. This corresponds to the intra-channel scattering between 
the chiral edge modes propagating in the same direction.
On the other hand, they are antiferromagnetically coupled between nearest neighbors, $x$ and $x+1$, via $J_j$. 
This term represents the inter-channel tunneling between the edge states in nearby channels, which propagate in opposite directions.
This model can be regarded as the $N_{\rm em}$ number of coupled $SU(2n)$ spin chains, where the index $j$ labels different spin chain.
Hence the $SU(2n)$ spins are aniferromagnetically coupled ($J_j$) within the same chain while the inter-chain 
coupling ($J_{jk}$) is ferromagnetic.
In this way, the quantum Hall transition maps onto a spin-Peierls transition of the spin system.  

Notice that the ferromagnetic coupling, between different chains for each $x$,
comes from the tunneling between the edge modes propagating in the same direction 
in the given channel. While the motion along the channel is ballistic, the tunneling
between the co-propagating chiral edge modes leads to diffusive motion in 
the transverse direction. This transverse diffusive mode is akin to the Goldstone
mode of ferromagnet. On the other hand, the anti-ferromagnetic coupling between nearby 
sites in each chain is the result of the tunneling between counter-propagating edge modes 
(allowing backscattering) in nearby channels. 
This anti-ferromagnetic coupling is the manifestation that ultimately the current is not separately 
conserved in each channel due to the coupling between counter-propagating edge modes 
in nearby channels. This is akin to non-conserved order parameter in antiferromagnets.

To gain some insight into this system, let us consider the case where the ferromagnetic couplings are large compared to the antiferromagnetic couplings. Here, we may view the $N_{\rm em}$-leg ladder as a single spin-$N_{\rm em}/2$ chain with anti-ferromagnetic nearest neighbor exchange 
(for example, in the case of $n=1$).  Crucially, since $N_{\rm em}$ is odd, a spin-Peierls transition still persists in this system and is in the same universality class as the spin-1/2 chain.  By contrast for $N_{\rm em}$ an even integer, as is the case for flux disorder {\it without} the zero energy Landau level, the resulting spin-chain remains in the Haldane gap phase, and the transition is lost!  Thus, the zero mode plays a key role not only in enabling a CF transition to occur, but also for the CF transition to be in the same universality class as the transition with $N_{\rm em} = 1$.   


For general $N_{\rm em}$ and $n$, it was argued in Ref. \onlinecite{PhysRevB.52.16646} that for large $N_{\rm em}$, the spins form a completely symmetric representation of $SU(2n)$ at each site. The inter-site couplings are still anti-ferromagnetic. 
In this case, the spin model is equivalent to the following non-linear sigma model:
\begin{gather}
\mathcal L = \frac{\pi \sigma^{\rm cf}_{xx}}{4}\mathrm{Tr}\left[\partial Q\right]^2 + \frac{N_{\rm em}}{16}\epsilon_{ij}\mathrm{Tr}\left[Q\partial_i Q \partial_j Q\right]
\end{gather}
where $Q = u^\dagger(\bm r) \Lambda u(\bm r)$ and  $u(\bm r) \in U(2n)$.
The symmetry of the non-linear sigma model is $U(2n)/U(n)\times U(n)$.

The second term is topological and $\int d^2 r\ \mathrm{Tr}\left[Q\partial_i Q \partial_j Q\right] = 16\pi i \ell$, where $\ell$ is an integer. Therefore, all values of $N_{\rm em}$ modulo 2 are equivalent. Since we have an odd number of chiral edge modes in general, we find that the composite fermions are described by:
\begin{gather}
\mathcal L = \frac{\pi \sigma^{\rm cf}_{xx}}{4}\mathrm{Tr}\left[\partial Q\right]^2 - \frac{1}{16}\epsilon_{ij}\mathrm{Tr}\left[Q\partial_i Q \partial_j Q\right]
\end{gather}
%
This suggests a duality in the class of the composite-fermion network models. Moreover, the integer quantum Hall transition described by the network models of composite fermions with an odd number of edge modes is equivalent to that of electrons. 
\section{Consequences of self-duality\label{sec:self_duality}}
In the previous 2 sections, we have provided two distinct lines of reasoning for the IQHIT to be self-dual (see Fig. \ref{fig:self_duality}). We now study its consequences. The self-duality imposes the following constraint on the conductivity tensors of electrons and composite fermions at the critical point:
\begin{gather}
\sigma_{ij} = \sigma_{ji}^{\rm cf}\label{self_duality_cond}
\end{gather}
In addition, the flux attachment procedure leads to the following relation between the corresponding resistivities:
\begin{gather}
\rho_{ij}^{\rm cf} = \rho_{ij} + 4\pi\epsilon_{ij}
\end{gather}

\begin{figure}
	\setlength{\unitlength}{0.1in} 
	\centering 
	\begin{picture}(28,16) 
	\linethickness{0.25mm}
	\put(0,14){\large 
		$\dfrac{\pi\sigma_{xx}}{4} \mathrm{Tr}\left[\partial Q\right]^2 + \dfrac{\pi |\sigma_{xy}|}{4} \epsilon_{ij}\mathrm{Tr}\left[Q\partial_iQ \partial_jQ\right]$
	}
	\put(12, 11.5){ \vector(0, -1){7} }
	\put(13.5, 8.5){Attach 2}
	\put(13, 7.2) {flux-quanta.}
	\put(0,1){\large $\dfrac{\pi\sigma_{xx}^{\rm cf}}{4} \mathrm{Tr}\left[\partial Q\right]^2 - \dfrac{\pi |\sigma_{xy}^{\rm cf}|}{4}  \epsilon_{ij}\mathrm{Tr}\left[Q\partial_iQ \partial_jQ\right]$
	}
	\end{picture}
	\caption{Self-duality of IQHIT in terms of the effective field theory. The theory at the top describes IQHIT at the half-filled Landau level. Upon attaching 2 flux quanta to the electron theory, the bottom theory of IQHIT in the CF-representation is obtained within mean-field approximation. Since it just changes the sign of the Hall conductivity, the critical point is self-dual.} 
	\label{fig:self_duality} 
\end{figure}

In terms of the conductance, i.e. $\sigma_{ij} = \left(\rho^{-1}\right)_{ij}$, it gives:
\begin{gather}
\sigma^{\rm cf}_{xx} = \frac{1}{16\pi^2}  \frac{\sigma_{xx}}{\left( \sigma_{xx} \right)^2 + \left( \sigma_{xy} - \frac{1}{4\pi} \right)^2}\label{dict_cond}
\end{gather}
From Eq. \eqref{self_duality_cond}, we get $\sigma_{xx} = \sigma_{xx}^{\rm cf}$. This alongside the constraint from particle-hole symmetry at $\nu=1/2$, i.e. $\sigma_{xy} = \frac{1}{4\pi}$, and Eq. \eqref{dict_cond} leads to:
\begin{gather}
\sigma_{xx}  = \frac{1}{4\pi}
\end{gather}
This is precisely the universal value of conductivity hinted at by both experiments and numerics.\cite{Huo1993, Wang1998} To emphasize the importance of this result, we note that while particle-hole symmetry fixes $\sigma_{xy} = \frac{1}{4\pi}$, there is no such symmetry argument for why $\sigma_{xx} = \frac{1}{4\pi}$. Its physical origin has been an open problem in the theory of quantum Hall transitions. Our work strongly suggests that such a universal conductivity arises from the presence of self-duality at the critical point under flux attachment.

\section{Discussion}
Previous work based on a composite boson description have also argued that the IQHIT is self-dual.  In the seminal work of Kivelson, Lee and Zhang (Ref. \onlinecite{Kivelson1992}) and others,\cite{Shahar1996, ShimshoniSondhiShahar1997} the IQHIT was mapped onto a composite-boson superconductor-insulator transition.  If the latter transition exhibited self-duality with a vanishing composite boson Hall conductance, the universal resistance at the IQHIT could be accounted for.  However, while such an argument seems plausible, the self-dual nature of the transition was never explicitly established. \footnote{In addition, classical studies of percolation also suggested a similar self-dual behavior.\cite{DykhneRuzin1994, RuzinShechao1995,Shimshoni2004}.} By contrast, here we have worked with composite fermions where self-duality can be derived at least within a mean-field approximation.  

The particular value of universal resistance, $\rho_{xx} = 2\pi$, has significance beyond the IQHIT. The law of corresponding states proposed in Ref. \onlinecite{Kivelson1992} implies that the resistivity tensor at all plateau-plateau transitions can be predicted using the universal resistivity tensor at IQHIT. In fact, experiments at $\nu=1/3$ to $\nu = 0$ transition have measured $\rho_{xx}\approx 2\pi$ consistent with the value found in this paper.\cite{Shahar1996} The extent to which superuniversality is manifested in our present framework, remains an interesting open problem for future work.

For the non-interacting description of electrons undergoing an IQHIT, there is only one energy at which the energy eigenstates are extended. As a consequence, at any finite temperature, the conductivity would remain zero at the critical point. \cite{Wang2000} The experimentally observed non-zero conductivity is usually explained by invoking inelastic scattering due to interactions. On the other hand, for composite fermions, we found that the states at all fermi-energies are extended at the critical point and therefore the conductivity is finite at non-zero temperatures. Therefore, we believe that the self-duality should be thought of as being a statement at the full disordered and interacting quantum critical point.

Let us mention some important caveats in our work. We have ignored the effects of gauge fluctuations on the composite fermions and the Coulomb interactions on the electrons. While the mean-field theory of composite fermions is valid for interactions longer ranged than Coulomb, it is unclear how the gauge fluctuations and disorder play out with each other in the presence of Coulomb interactions.  In addition, the Coulomb interactions in the electron theory can change the universality class of electron's critical point. Experiments have found that the dynamical exponent takes the value $z=1$ at $\nu=1 \rightarrow 0$ transition. This is inconsistent with the prediction of non-interacting electron theory, i.e. $z=2$, due to the finite density of states at the transition. While this discrepancy has been alluded to the effect of long-range Coulomb interaction in literature\cite{Polyakov1998,Huckestein1999}, it still remains as an open question.
The extent to which these caveats affect the self-duality of integer quantum Hall transition 
is an outstanding question for future studies.  The composite fermion approach studied here appears to have considerable promise in addressing these open questions.  It is possible that such theories may also have implications for related transitions, such as the magnetic field tuned superconductor to insulator transition; considerable experimental evidence points to the self-dual nature of this transition.\cite{Paalanen1992,Breznay2016}  While some preliminary theoretical work has been done in this direction\cite{MulliganRaghuCFsatSIT, PhysRevB.95.045118}, much remains to be explored. 

%

\acknowledgments
The authors thank D. Arovas, Jing-Yuan Chen, H. Goldman, P. Goswami, S. Kivelson, R. Laughlin, R. Shankar, and especially M. Mulligan for illuminating discussions.  We also thank M. Mulligan for collaboration on related topics.  
P.K. and  S.R are supported in part by the 
the DOE Office of Basic Energy Sciences, contract DEAC02-76SF00515.
Y.B.K. is supported by the NSERC of Canada and the Killam Research Fellowship of the Canada Council for the Arts.

\appendix
\section{Non-linear sigma model from 2d nonabelian anomaly\label{topological_term_derivation}}
In this appendix, we explicitly derive the non-linear sigma model for Dirac fermions with vector potential disorder in 2+1 dimensions. We will find a kinetic energy term for the diffusive modes representing the longitudinal conductivity and a topological theta term  representing the physics of Hall conductivity.  The latter term reflects the fact that the Dirac fermions are at a critical point between two integer quantum Hall states.  We derive the topological term explicitly by making use of chiral rotations, which are anomalous, and lead to purely imaginary Jacobians of the path integral measure.\cite{Fujikawa1979, Fujikawa1980}  

The theory consists of a single, two-component Dirac fermion in $d=2+1$ spacetime dimensions in the presence of a random vector potential, and a non-zero uniform chemical potential.  The Hamiltonian in two-dimensional space in the presence of disorder is:
\begin{gather}
H = H_0 + H_{dis}\\
H_0 = \int d^2 x  \  \psi^{\dagger}  \left[    -i \tau_j   \partial_j      - \mu \right]\psi\\
H_{dis} = \int d^2 x  \   a_j \psi^{\dagger}  \tau_j \psi
\end{gather}
where $j=1,2$, $\tau_j = (\sigma^x, \sigma^y)$ are the Pauli-matrices. For convenience, let's also define $\tau^0 = \mathbb{1}_2$ and $\tau_3 = \sigma^z$. The spatial components of the vector potential ${\bf a}$,  are quenched random variables chosen from the following probability distribution
\begin{equation}
P[a] = \frac{1}{\mathcal N} e^{-\frac{1}{2g} \int d^2x\ a_j^2}
\end{equation}

This implies that (defining the random magnetic field $b = \epsilon_{ ij} a_i a_j$ )
\begin{gather}
\overline{a_\mu(\bm k)a_\nu(\bm k')} = (2\pi)^2g \delta_{\mu\nu}\delta^2(\bm k + \bm k')\\
\overline{b(\bm k )b(\bm k')} = (2\pi)^2 \Delta k^2 \delta^2(\bm k + \bm k').
\end{gather}
Note that there is zero uniform ($k=0$) piece to $b$. The Grassmannian manifold arises when studying the conductivity of this system.  We will only be interested in the dc limit in what follows, and the dc conductivity involves  disorder averaged products of retarded and advanced Green functions evaluated at the chemical potential. In a functional integral approach, the generator of such correlation functions is:
\begin{align}
Z &=  \int \prod_{ \sigma = \pm 1} D  \psi^{\dagger}_{ \sigma} D \psi_{ \sigma}\ e^{-S}, \nonumber \\
S &= \int d^2 x \left\{ \sum_{ \sigma,  \sigma' = \pm 1}   \psi^{\dagger}_{ \sigma} \left[ \left(E   - H_0 \right) \delta_{ \sigma  \sigma'}        + i \eta  \sigma^z_{ \sigma  \sigma'} \right] \psi_{  \sigma'} \right. \nonumber\\
&\ \ \ \ \ \ \ \ \ \ \ \ \ \ \ \ \ \ \ \ \ \ -   \left. a_j  \cdot \sum_{ \sigma = \pm 1} \psi^{\dagger}_{ \sigma}  \tau_j \psi_{ \sigma} \right\}
\end{align}
The discrete index $ \sigma = \pm 1 $ labels retarded and advanced correlation functions, $E$ is the energy above the Fermi level, and $\eta$ is a positive infinitesimal.  
Observe that since interactions are ignored, the action arises only from the spatial components and time is no longer present: the theory to be derived ``lives" in one lower dimension than the original degrees of freedom.  From this point onwards, it will be helpful to think of this theory as living entirely in $2+0$ dimensions.  

The disorder averaging over the quenched random vector potentials is done using the replica trick.  We introduce $n$ copies of fermions of both retarded and advanced species, and performing the disorder average, we obtain

\begin{widetext}
\begin{gather}
\overline{Z^n} =    \int \prod_{ \sigma = \pm 1} \prod_{\alpha = 1}^nD  \psi^{\dagger}_{  \sigma \alpha} D \psi_{  \sigma \alpha}\ e^{-S_1 - S_2},   \nonumber \\
S_1 = \int d^2 x   \sum_{ \sigma,  \sigma' = \pm 1} \sum_{\alpha=1}^n  \psi^{\dagger}_{ \sigma \alpha} \left[ \left(E + i \tau_j   \partial_j  \right) \delta_{ \sigma  \sigma'} +  i \eta  \sigma^z_{ \sigma  \sigma'} \right] \psi_{ \sigma' \alpha}    \nonumber \\
\begin{split}
S_2 &= -\frac{\Delta}{2} \int d^2 x \sum_{ \sigma,  \sigma' = \pm 1} \sum_{\alpha,\beta=1}^n \left( \psi^{\dagger}_{ \sigma \alpha}  \tau_j \psi_{  \sigma \alpha} \right) \left( \psi^{\dagger}_{i  \sigma' \beta} \tau_j \psi_{i  \sigma' \beta} \right) \\
&\equiv  -\frac{\Delta}{2} \int d^2 x \sum_{ \sigma,  \sigma' = \pm 1} \sum_{\alpha,\beta=1}^n \bm j_{  \sigma \alpha} \cdot \bm j_{  \sigma' \beta}
\end{split}
\end{gather}
\end{widetext}
we have absorbed $\mu$ inside $E$ by redefining $E \rightarrow E -\mu$. First we treat the interaction between different replicas at saddle point level. We seek an order parameter that is
off-diagonal in retarded/advanced space. For this it proves useful to make use of the following Pauli-matrix identity:
\begin{align}
- \frac{\Delta}{2} \bm j_{  \sigma \alpha} \cdot \bm j_{  \sigma' \beta} &= -\frac{\Delta}{2}  \left( \psi^{\dagger}_{  \sigma \alpha } \tau^3 \psi_{  \sigma' \beta} \right)   \left( \psi^{\dagger}_{  \sigma' \beta } \tau^3  \psi_{ \sigma \alpha} \right)  \nonumber\\
&\ \ \ \ + \frac{\Delta}{2}  \left( \psi^{\dagger}_{  \sigma \alpha } \tau^0 \psi_{  \sigma' \beta } \right) \left(  \psi^{\dagger}_{  \sigma' \beta} \tau^0 \psi_{  \sigma \alpha } \right)
\end{align}
Decoupling the first term through a Hubbard-Stratonovic transformation will correspond to the diffusion of the z-component of the spin. Since, the Dirac Hamiltonian breaks spin-conservation, it does not diffuse in space. Thus, we can ignore this disorder term. So, let's decouple only the second term via the identity:
\begin{align}
e^{-\frac{\Delta}{2}  \left( \psi^{\dagger}_{  \sigma \alpha } \tau^0 \psi_{  \sigma' \beta } \right) \left(  \psi^{\dagger}_{  \sigma' \beta} \tau^0 \psi_{  \sigma \alpha } \right)} &\propto \nonumber\\
   \int d Q_{\alpha\beta;  \sigma  \sigma'} &\ e^{-\left[ \kappa {\rm Tr}\left(Q^\dagger Q \right) + i \gamma Q_{\alpha\beta;  \sigma  \sigma'} \psi^{\dagger}_{ \alpha \sigma} \tau^0 \psi_{ \beta  \sigma'}  \right]}
\end{align}
where $\kappa = \frac{\pi \rho_F}{4\tau}, \gamma = \frac{1}{2\tau}, \frac{1}{\tau} = 2\pi \Delta \rho_F$ and $\rho_F$ is the density of states of the Dirac fermion at the chemical potential. We have chosen $Q$ to be Hermitian, i.e. $Q^\dagger = Q$.

The full action reads:
\begin{align}
S^{(rep)} &= \int d^2x\ \psi^\dagger\left[E + i\tau_j\partial_j  + i\eta {\Lambda^z} + i\gamma Q\right]\psi\nonumber\\
&\ \ \ \ \ \ \ \ \ \ \  + \kappa\int d^2x\ {\rm Tr}[Q^\dagger Q] \label{action_psi_Q}
\end{align}
where we have suppressed the retarded/advanced and replica indices for brevity.

$Q$-fields are the long wavelength degrees of freedom for this theory. So, let's integrate out fermions and find the effective action in terms of $Q$:
\begin{align}
S_{\rm eff}[Q] &= \int d^2 x\ \left[\kappa {\rm Tr}[Q^\dagger Q ]\right. \nonumber\\
- &\left.{\rm Tr}\ln\left[E+i\tau_j \partial_j+i\eta{\Lambda^z} + i\gamma Q\right]\right]\label{action_Q}
\end{align}
where ${\Lambda^z} \equiv \sigma^z_{ \sigma \sigma'}\delta_{\alpha\beta}$. The saddle point equation from varying the action with respect to $Q$ is:
\begin{gather}
\pi \rho_F Q = {\rm Tr}_\tau \left[i\int \frac{d^2q}{(2\pi)^2} \frac{1}{E -\slashed {q} + i\eta {\Lambda^z} + i\gamma Q}\right]
\end{gather}
where ${\rm Tr}_\tau$ represents trace over $\tau$-matrix indices. We have assumed a translationally-invariant replica-diagonal saddle.
The integrals can be done by introducing the density of states and integrating over energy. Then we find the saddle point solution to be:
\begin{gather}
Q = {\Lambda^z}
\end{gather}
Except for the infinitesimal $\eta$ term, the action in Eq. \eqref{action_psi_Q} is invariant under unitary transformation of the fermion fields in replica and retarded/advanced spaces. Thus, it possesses $U(2n)$ symmetry. The saddle point solution, that's chosen by the infinitesimal $\eta$ term, breaks it down to $U(n)\times  U(n)$. Under a global unitary transformation of the fermions $\psi \rightarrow u \psi$, $Q$ transforms as $Q \rightarrow u^\dagger Q u$. Thus, the non-linear sigma model for diffusion can be obtained by constraining $Q$ matrices as:
\begin{gather}
Q^2 = \mathbb{1}_{2n}\nonumber\\
{\rm Tr}[Q] = 0
\end{gather}
The coset space of this NLSM is $U(2n)/U(n)\times U(n)$, i.e. the unitary class. This is expected based on the fact that the Dirac fermion theory breaks time-reversal and spin-conservation. Thus, the $Q$-fields describe the diffusion of just the fermion density.

Since this theory involves a single Dirac fermion, the effects of chiral anomaly need to be analyzed carefully. For example, it is known that in half-filled 1+1 dimensional Hubbard model, it leads to a topological term that describes the Berry phase effects for SDW modes.\cite{Nagaosa1996} In our case, as we'll see, it will give rise to a topological term describing the physics of Hall conductivity and integer quantum Hall to insulator transition. 

The fermion part of the action in Eq. \eqref{action_psi_Q} is:
\begin{align}
S &= \int d^2x\ \psi^\dagger\left[E + i\tau_j\partial_j  + i\gamma Q\right]\psi
\end{align}
Before proceeding, we observe that we may interpret this 2+0 dimension theory independently of the parent 2+1 dimensional theory. We view it as 2-dimensional Dirac fermions in Euclidean space. Let's take $\gamma^0= \tau^1$ and $\gamma^1 = \tau^2$ so that the Euclidean signature is $(+,+)$. Quite conveniently, we can interpret $\psi^\dagger \equiv i\bar\psi$ since the gamma matrices satisfy $\{\gamma^\mu, \gamma^\nu\} = 2\delta^{\mu\nu}$. The action now reads:
\begin{align}
S &= -\int d^2x\ \bar\psi\left[-iE + \slashed\partial  + \gamma Q\right]\psi
\end{align}
Using $\bar\psi \equiv\psi^\dagger \gamma^0 $, we arrive at:
\begin{align}
S &= -\int d^2x\ \psi^\dagger\left[-iE\tau^1 + \partial_1 + i\tau^3\partial_2   + \gamma \tau^1 Q\right]\psi
\end{align}
Let's now go to a rotating frame of reference via the substitution $Q = u{\Lambda^z} u^\dagger$ so that $Q$ is always aligned along ${\Lambda^z}$ locally in space and the fermions experience a gauge field in exchange. $u \in U(2n)$ is a slowly varying matrix. To do this, we transform $\psi= u \tilde\psi$ and $\psi^\dagger = \tilde\psi^\dagger u^\dagger$,
\begin{align}
S &= -\int d^2x\ \tilde\psi^\dagger \begin{bmatrix}
\partial_+ + u^\dagger \partial_+ u & -i E + \gamma {\Lambda^z} \\ -i E + \gamma {\Lambda^z} & \partial_- + u^\dagger \partial_- u
\end{bmatrix}\tilde\psi\label{action_gauge}
\end{align}
where we have explicitly written out the $\tau$-matrix components of the action and $\partial_{\pm} = \partial_1 \pm i \partial_2$. The Fermi energy has become an imaginary mass for the 2-dimensional Dirac fermions. Notice that the field strength of the $U(2n)$ gauge field $a_\mu \equiv iu^\dagger\partial_\mu u$ satisfies:
\begin{gather}
f_{\mu\nu} = i[D_\mu,D_\nu] = 0.\label{field_strength_a}
\end{gather}
This should not be surprising because the gauge field was introduced through a gauge transformation of fermion fields. However, $a_\mu$ is not unphysical. The action spontaneously breaks $U(2n)$ symmetry to $U(n)\times U(n)$ and thus some of the components of $a_\mu$ correspond to the resulting Goldstone modes. This is made clear by the following relation:
\begin{gather}
\partial_\mu Q = iu[{\Lambda^z},a_\mu]u^\dagger\label{goldstone_gauge}
\end{gather}
All components of $a_\mu$ that commute with ${\Lambda^z}$ are unbroken gauge degrees of freedom while the remaining ones are Goldstone modes.

For chiral anomaly, one should analyze the full action containing the unbroken non-abelian gauge symmetry $U(n)\times U(n)$. However, the same result can be achieved by treating each replica separately and adding up their anomaly contributions. Such contributions can be calculated using the formalism of just the abelian $U(1)$ anomalies. For the reasons of simplicity and the pedagogical value, we do the latter first in section \ref{sec:u1_anomaly}. For a more technically complete analysis, we direct the reader to section \ref{sec:unun_anomaly}.


\subsection{Topological term using $U(1)$ anomalies of replicas\label{sec:u1_anomaly}}

In this subsection, we will look at each replica separately and find the topological term that results from adding up their contributions. Thus, we split \footnote{Within a single replica, there are two $U(1)$ gauge fields. One of them is the $\Lambda^a$-component of $a_\mu$, the other corresponds to $\mathbb{1}^a  \equiv \delta_{\sigma\sigma'}\delta_{\alpha\beta}\delta_{\alpha a} $ component. However, we can work in a gauge where the latter are zero.} $a_\mu = \mathcal{A}^a_\mu {\Lambda^a} + \delta a_\mu$, where ``$a$'' is a replica index, $\mathcal{A}^a_\mu$ is a set of $n$ abelian $U(1)$ gauge fields and:
\begin{gather}
\Lambda^a_{\alpha\beta;\sigma\sigma'} = \sigma^{z}_{\sigma\sigma'}\delta_{\alpha\beta}\delta_{\alpha a}\\
\mathcal{A}^a_\mu = \frac{1}{2} {\rm Tr}[{\Lambda^a} a_\mu]\\
F_{\mu\nu}^a = \partial_\mu \mathcal{A}^a_\nu - \partial_\nu \mathcal{A}^a_\mu
\end{gather}
Thus, we are looking at ${U(1)}^n$ part of the unbroken gauge symmetry $U(n)\times U(n)$ that ignores mixing between different replicas. 

We will treat $\delta a_\mu$ as a perturbation. So, let's set it to equal zero and write the action as:
\begin{gather}
S[\mathcal A] = -\int d^2x\ \sum_{a=1}^n\tilde\psi^\dagger_a \begin{bmatrix}
\partial_+ -i \mathcal{A}^a_+  & -i E + \gamma {\sigma^z} \\ -i E + \gamma {\sigma^z} & \partial_- -i \mathcal{A}_-^a 
\end{bmatrix}\tilde\psi_a 
\end{gather}
where $\tilde \psi_a, \tilde \psi_a^\dagger$ represent the Dirac field corresponding to replica index ``$a$''.  

This action is similar in structure to the $SU(2)/U(1)$ theory in Ref. \onlinecite{Nagaosa1996}. So, we will closely follow their approach for deriving the topological term.
We can interpret the above action as two Dirac fermions (retarded and advanced) that have masses related by $m_+ = -m_-^* = -iE + \gamma$. They also have opposite charges under the $U(1)$ gauge field $\mathcal{A}^a_\mu$. Thus, we can write the partition function as:
\begin{gather}
Z[\mathcal{A}] = \prod_{a=1}^n Z[\mathcal{A}^a,m]Z[-\mathcal{A}^a, -m^*]\\
Z[\mathcal{A}^a,m] = \int \mathcal{D}\bar\psi_a \mathcal{D}\psi_a \ e^{-S[\mathcal{A}^a,m]}\nonumber\\
S[\mathcal{A}^a,m] = -\int d^2x\ \psi^\dagger_a \begin{bmatrix}
\partial_+ -i \mathcal{A}_+^a & m \\ m & \partial_-^a -i \mathcal{A}_-^a 
\end{bmatrix}\psi_a
\end{gather}
where $m = -i E  + \gamma$ and $\psi_a,\psi^\dagger_a$ are now $2$-dimensional Grassmann variables. 

We will now find the relative phase between the retarded and advanced Dirac fermion actions. To do this, we can first reverse the sign of the mass of advanced Dirac fermions through a chiral rotation: $\psi_a \rightarrow e^{i\tau^3 \pi/2}\psi_a, \psi^\dagger_a \rightarrow \psi^\dagger_a e^{-i\tau^3 \pi/2}$. Using Fujikawa's method, the Jacobian of this transformation gives us the topological term:\footnote{One should find the Jacobian of the finite chiral transformation $U_{ch} = e^{i\tau^3\pi/2}$ by building it up continuously from infinitesimal chiral rotations of the form $U_{ch}(\alpha) = e^{i\alpha \tau^3 \pi/2}$. However, in our case, $\gamma^\mu D_\mu$ is independent of $\alpha$ and thus the calculation simplifies.}
\begin{gather}
Z[-\mathcal{A}^a, -m^*] = e^{-S_{top}^a}Z[-\mathcal{A}^a, m^*]\nonumber\\
S_{\rm top}^a =  \frac{i}{4} \int d^2x\ \epsilon^{\mu\nu}F^a_{\mu\nu}\\
\begin{split}
S_{\rm top} &= \sum_{a=1}^n S_{\rm top}^a\\
&= \frac{i}{4} \int d^2x\ \epsilon^{\mu\nu} \sum_{a=1}^n F^a_{\mu\nu}
\end{split}
\end{gather}
Quite nicely, the topological term is a sum over the $U(1)$ abelian anomaly of each individual replica. We can convert it in terms of the $Q$-matrix using the following identity:
\begin{gather}
\begin{split}
\sum_{a=1}^n F^a_{\mu\nu} &= \frac{1}{2}{\rm Tr} \left[\left(\partial_\mu a_\nu - \partial_\mu a_\nu\right)\sum_a\Lambda^a\right]\\
&= \frac{1}{2}{\rm Tr} \left[\left(\partial_\mu a_\nu - \partial_\mu a_\nu\right)\Lambda^z\right]\\
&= \frac{i}{4} {\rm Tr}\left[Q\partial_\mu Q \partial_\nu Q\right]
\end{split}\label{F_replica}\\
S_{\rm top} = -\frac{1}{16} \int d^2x\ \epsilon^{\mu\nu} {\rm Tr}\left[Q\partial_\mu Q \partial_\nu Q\right]
\end{gather}

Further, we can use the charge and Hermitian conjugation properties of 2d Dirac fermions to arrive at:\cite{ZinnJustin1996}
\begin{align}
Z[\mathcal{A}] &= e^{-S_{top}} \prod_{a=1}^n Z[\mathcal{A}^a,m]Z[-\mathcal{A}^a, m^*]\\
&= e^{-S_{top}} \prod_{a=1}^n |Z[\mathcal{A}^a,m]|^2
\end{align}
Thus, the topological term is the imaginary part of the action that results from anomalous chiral rotation between Dirac fermions of the retarded and advanced kind.

\subsection{Topological term using the full action\label{sec:unun_anomaly}}
In this section, we will analyze the chiral anomaly keeping all gauge field components. For this purpose, we'll generalize the approach of Ref. \onlinecite{Goswami2011} to $U(2n)/U(n)\times U(n)$ case.

It will be useful to split the gauge field $a_\mu = iu^\dagger \partial_\mu u$ into two parts. First, $C_\mu$ are the gauge field components that correspond to the  unbroken gauge symmetry $U(n)\times U(n)$ and second, $G_\mu$ are the remaining gauge fields. We can show the following relations:
\begin{gather}
a_\mu = C_\mu + G_\mu\nn\\
[C_\mu,\Lambda^z] = 0\nn\\
\{G_\mu,\Lambda^z\}= 0\nn\\
C_\mu = \frac{1}{2}\left(a_\mu + \Lambda^z a_\mu \Lambda^z\right)\\
G_\mu = \frac{1}{2}\left(a_\mu - \Lambda^z a_\mu \Lambda^z\right)
\end{gather}
$C_\mu$ are block diagonal $2n\times 2n$ matrices that are diagonal in retarded/advanced space. On the other hand, $G_\mu$ are off-diagonal in retarded/advanced space. Notice that only $G_\mu$ correspond to Goldsone modes as one can see from Eq. \eqref{goldstone_gauge}. This is related to the fact that $U(2n)$ gauge symmetry is broken down to $U(n)\times U(n)$ and $C_\mu$ correspond to the latter. Further, although the field strength of $a_\mu$ is zero (Eq. \eqref{field_strength_a}), the field strength of $C_\mu$ is:
\begin{align}
f^C_{\mu\nu} &= \partial_\mu C_\nu - \partial_\nu C_\mu - i [C_\mu, C_\nu]\nonumber\\
&= \frac{i}{4} u \left(Q\partial_{[\mu}Q \partial_{\nu]}Q\right)\Lambda^z u^\dagger
\end{align}
where $[\mu\cdots \nu]$ means anti-symmetrization with respect to $\mu,\nu$. Clearly, $C_\mu$ can not be completely gauged away. The topological term can be formed by two possible combinations of the gauge fields:
\begin{align}
{\rm Tr}\left[Q\partial_\mu Q \partial_\nu Q\right] &= -2i{\rm Tr}[\Lambda^z f^C_{\mu\nu}] \\
&= 2{\rm Tr}\left[\Lambda^z \left[G_\mu, G_\nu\right]\right]\label{G_top}
\end{align}

We can rewrite the action in Eq. \eqref{action_gauge} as:
\begin{gather}
S = -\int d^2x\ \overline{\tilde\psi} \left[\slashed\partial -i\slashed C - i\slashed G-iE + \gamma\Lambda^z\right]\tilde\psi
\end{gather}

We'll now perform the chiral rotation $U_{ch} = e^{i \gamma^5 \left(\Lambda^z - \mathbb{1}\right)\pi/4}$ on this theory as we did in subsection \ref{sec:u1_anomaly} (where $\gamma^5 \equiv i\tau^1\tau^2= -\tau^3$). This makes the masses of retarded and advanced fermions become complex conjugates of each other. The gauge invariant regularization for the anomaly calculation involves $\slashed D \equiv \gamma^\mu \left(\partial_\mu - ia_\mu\right)$ which, unlike the previous subsection, changes under this operation.  So, the finite chiral rotation needs to be built up from successive infinitesimal rotations. To do so, we define the continuous chiral rotation $U_{ch}(\alpha)=e^{i\alpha \gamma^5 \left(\Lambda^z - \mathbb{1}\right)\pi/4}$. The fermion fields transform as: $\tilde\psi\rightarrow U_{ch}(\alpha)\tilde\psi$ and $\overline{\tilde\psi} \rightarrow \overline{\tilde\psi} U_{ch}(\alpha)$. $\slashed D$ transforms to:
\begin{gather}
\slashed D_\alpha = \slashed \partial -i \slashed C - i\left[\cos\left(\frac{\alpha \pi}{2}\right) + i\gamma^5 \Lambda^z\sin\left(\frac{\alpha \pi}{2}\right)\right] \slashed G
\end{gather}
The Jacobian of the finite chiral rotation is:
\begin{gather}
J = \exp \left[\frac{i}{2\pi} \frac{\pi}{4} \int_0^1 d\alpha\ {\rm Tr}\left[\left(\Lambda^z-\mathbb{1} \right) \gamma^5 \slashed D_\alpha^2\right]\right]
\end{gather}
There are two terms that are proportional to the topological term:
\begin{gather}
J = \exp \left[\frac{i}{2\pi} \frac{\pi}{4}\int_0^1 d\alpha \int d^2x\ \left(T_1+T_2\right)\right] \nn\\
T_1 = -\epsilon^{\mu\nu} {\rm Tr}\left[\Lambda^z f^C_{\mu\nu} \right]\\
T_2 = i\epsilon^{\mu\nu} \cos \alpha\pi \ {\rm Tr}\left[ \Lambda^z \left[G^\mu, G_\nu\right]\right] 
\end{gather}
$T_2$ integrates out to zero and $T_1$ gives the topological term. This expression makes it clear why treating each replica separately worked in the subsection \ref{sec:u1_anomaly}. Due to the appearance of trace in the $T_1$ term, only the $\Lambda^z$-component of the $U(n)\times U(n)$ field strength $f^C_{\mu\nu}$ matters. This is exactly equal to what we got in Eq. \eqref{F_replica}. We obtain the Jacobian:
\begin{gather}
J = \exp \left[\frac{\epsilon^{\mu\nu}}{16} \int d^2x\ {\rm Tr}\left[Q\partial_\mu Q \partial_\nu Q\right]\right]
\end{gather}
The topological term in the action is:
\begin{gather}
S_{top} = -\frac{\epsilon^{\mu\nu}}{16} \int d^2x\ {\rm Tr}\left[Q\partial_\mu Q \partial_\nu Q\right]
\end{gather}

\subsection{Kinetic energy term for Goldstone modes}
Lastly, we can find the kinetic energy term by integrating out fermions and expanding the action in Eq. \eqref{action_gauge} to second order in $a_\mu$. It is given by:
\begin{gather}
S_{\rm eff}^{(2)} = \frac{\pi\sigma_{xx}}{4}\int d^2x\ {\rm Tr}[\partial Q]^2
\end{gather}
where $\sigma_{xx}= \frac{1}{4\pi \Delta}$. The full action for the NLSM now reads:
\begin{gather}
S_{\rm eff}[Q] = \int d^2x\ \left[\frac{\pi\sigma_{xx}}{4}{\rm Tr}[\partial Q]^2-\frac{\epsilon^{\mu\nu}}{16} {\rm Tr}[Q\partial_\mu Q \partial_\nu Q]\right]
\end{gather}
%


\subsection{Relation between $g=2$ and Dirac-fermion theory}

In this subsection, we point out some minor differences between the Dirac theory and $g=2$ theory. In section \ref{sec:NLSM}, we had mentioned that $g=2$ theory is slightly different from the Dirac theory in the sense that one of the components of $\Psi$ field doesn't have a time-derivative. This doesn't affect the static physical observables like the dc conductivities, however, leads to changes in the definitions of other quantities. To see this, we utilize the  the relation between $g=2$ and Dirac fermion Green functions:\cite{Takahashi2002}
\begin{gather}
G_{R,A}^{g=2}(\mu^{g=2}) = \frac{m}{\mu}\left(G_{R,A}(\mu) - G_{A,R}(-\mu)\right)_{\downarrow\downarrow}\nonumber\\
\mu^{g=2} = \frac{\mu^2}{2m}\nonumber\\
G_{R,A}(\mu) = \frac{1}{\mu - \bm\sigma.\bm p \pm \frac{i}{2\tau}}
\end{gather}
where ``$\downarrow\downarrow$'' means the down spin component. This gives:
\begin{gather}
G_{R,A}^{g=2}(\mu^{g=2}) = \frac{1}{\mu^{g=2} - \frac{p^2}{2m} \pm i\frac{\mu}{2m\tau}}
\end{gather}
Using this and $\rho^{g=2}_F = \frac{m}{2\pi},\ \rho_F = \frac{\mu}{2\pi}$; we get
\begin{gather}
\frac{1}{\tau^{g=2}} = \frac{\mu}{m\tau} = 2\pi g v_F^2 \rho_F^{g=2}\nonumber\\
D^{g=2} = \frac{v_F^2 \tau^{g=2}}{2} = \frac{1}{4\pi g \rho^{g=2}_F}\nonumber\\
\sigma^{g=2}_{xx} = D^{g=2} \rho^{g=2}_F = \frac{1}{4\pi \Delta}
\end{gather}
where $D^{g=2}$ is the diffusion constant and $v_F = \frac{k_F}{m}$ is the Fermi velocity. The longitudinal conductivity for $g=2$ is same as what we got for the Dirac fermion problem. In addition, the topological term stays unchanged because it was derived using the chiral anomaly of 2+0 dimensional Dirac fermions.


\section{IQHIT self-duality from the Dirac theory of composite fermions\label{ff_duality}}
We have presented the self-duality at IQHIT taking the viewpoint of the HLR theory. However, Son has conjectured a Dirac fermion description\cite{Son2015} of the half-filled Landau level that has a manifest particle-hole symmetry. In this appendix, we present the self-duality again, taking the latter perspective and find that the description stays exactly the same. In fact, we'll discover that the mean-field descriptions of HLR and Son's theories with disorder are physically and mathematically close. This treatment makes the self-duality a general feature of IQHIT, not tied to a specific composite fermion theory.

Let's consider 2+1 dimensional Dirac fermions at zero chemical potential in the presence of a large magnetic field. 
\begin{gather}
Z[A] = \int D \bar \Psi D \Psi\ e^{i \int \mathscr{L}[A]},\\ 
\mathscr{L}[A]= i\bar\Psi\slashed D_A\Psi + \frac{1}{8\pi}AdA + \cdots \label{Dirac_electron}
\end{gather}
where $i\slashed D_A = \gamma^\mu (\partial_\mu -iA_\mu)$, $\nabla \times \bm A = B$, and the second term represents the contribution from a massive partner Dirac fermion. Since the Dirac fermion is at zero chemical potential, it corresponds to the half-filled zeroth Landau level of the Dirac fermion. Son proposed the following composite fermion theory:
\begin{gather}
Z[A] = \int D \bar \psi D \psi Da\ e^{i \int \mathscr{L}_{\rm cf}[A]}\\
\mathscr{L}_{\rm cf}[A] = i\bar\psi\slashed D_a\psi  -\frac{1}{4\pi}Ada + \frac{1}{8\pi}AdA + \cdots\label{Dirac_cf}
\end{gather}
\begin{figure}
	\centering
	\includegraphics[width=2.in]{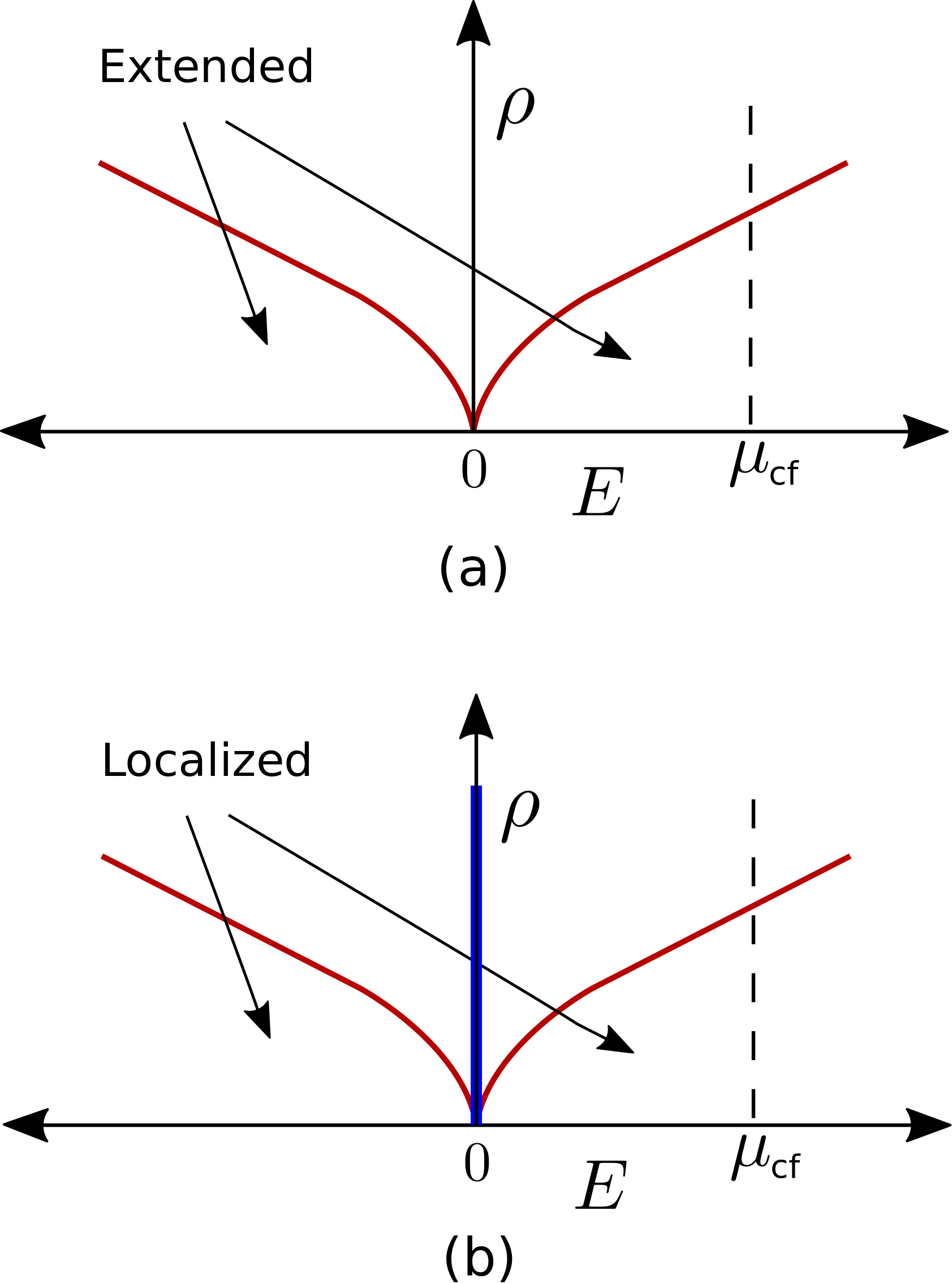}
	\caption{A schematic for density of states of the Dirac fermion Hamiltonian $\mathcal{H}^D_{\rm cf}$ in Eq. \eqref{Dirac_Hamiltonian} for (a) $b_0 = 0\ $ \cite{PhysRevB.50.7526} and (b) $b_0\neq 0$. All states are extended for $b_0 = 0$ due to the presence of statistical time-reversal symmetry. For $b_0 \neq 0$, there are zero-energy states that contribute a Hall conductance of $\sigma_{xy}^{\rm cf} = \frac{1}{4\pi}\mathrm{sgn}[b_0]$. All positive energy extended states are assumed to have levitated up. Therefore, $b_0=0$ is the critical point between the two integer quantum Hall states.}
	\label{fig:dirac_dos}
\end{figure}

These composite fermions are at a finite chemical potential $\mu_{\rm cf}$ so that $\rho_{\rm cf}\bar\psi\gamma^0\psi = \frac{B}{4\pi}$.
Analogous to what we did for the HLR composite fermions, let's add chemical potential disorder to this theory and construct the mean-field Hamiltonian for composite fermions. This can be done by shifting $A_0 \rightarrow A_0 + V(\bm r)$. The relevant term in the Lagrangian is:
\begin{gather}
\mathscr{L}_{\rm dis} = -\frac{1}{4\pi}V(\bm r)b
\end{gather}
It acts as a source of effective magnetic field $b \equiv \nabla \times \bm a$. Therefore, the Dirac composite fermions experience a random effective magnetic field due to a potential disorder in the electron coordinates. For a smooth disorder, the random magnetic field will be proportional to $V(\bm r)$ as we saw in the HLR theory, Eq. \eqref{slave_disorder_relation}. Using these, we obtain the mean-field Hamiltonian of the Dirac composite fermions:
\begin{gather}
\mathcal{H}^D_{\rm cf} = \left(\bm p - \bm a(\bm r)\right).\bm\sigma\label{Dirac_Hamiltonian}
\end{gather}
where $\bm a(\bm r)$ is a quenched vector-potential representing the random magnetic field and $\bm \sigma = (\sigma^x, \sigma^y)$ are Pauli matrices. 

It now becomes clear that the physics of the Dirac composite fermion theory is identical to that of $g=2$ theory of composite fermions in Eq. \eqref{HLR_Hamiltonian}. In section \ref{sec:NLSM}, we had mapped the $g=2$ mean-field Lagrangian to a Dirac fermion with random magnetic field, i.e. precisely the mean-field Hamiltonian obtained in Son's theory of composite fermions. 
%
%

Let's now discuss the appearance of IQHIT self-duality in the Dirac fermion language. 
The parent Dirac ``electron'' theory in Eq. \eqref{Dirac_electron} has an IQHT in the presence of chemical potential disorder. 
This can be seen by tuning the chemical potential $\mu$ across $\mu=0$. 
The random chemical potential broadens the zeroth Landau level. 
It has an extended state at zero energy if the odd moments of the disorder vanish, i.e. the disorder is statistically particle-hole symmetric. 
Also, all other states are localized. 
Since a full zeroth Landau level has a Hall conductance $\sigma_{xy} = \frac{1}{4\pi}$, the theory exhibits the following phase transition at $\mu=0$:
\begin{gather}
\sigma_{xy} = \left\{\begin{matrix}
\frac{1}{4\pi}, & \mu > 0 \\ -\frac{1}{4\pi}, & \mu < 0
\end{matrix}\right.
\end{gather}
Let's now discuss its dual interpretation in the Dirac composite fermion language using the Hamiltonian in Eq. \eqref{Dirac_Hamiltonian}. Since, the chemical potential $\mu$ is a source of effective magnetic field for the composite fermion, tuning across the half-filled Landau level corresponds to tuning the effective magnetic field across zero. So, as we did for HLR theory, let's split $b(\bm r) = b_0 + \tilde b(\bm r)$, where $b_0 = \frac{1}{L^2}\int d^2r\ b(\bm r)$ is the uniform part and $\tilde b(\bm r)$ is the quenched disorder part of the effective magnetic field.

The density of states for $b_0 \neq 0$ and $b_0 = 0$ is plotted in Fig. \ref{fig:dirac_dos}. For $b_0\neq 0$, the Dirac fermion has exact zero modes\cite{PhysRevA.19.2461} analogous to the lowest Landau level. A filled set of such zero-modes contributes a Hall conductivity of $\sigma^{\rm cf}_{xy} = \frac{1}{4\pi}{\rm sgn}[b_0]$. Moreover, all positive energy states are localized. Thus, we find that $b_0=0$ is a critical point between the following integer quantum Hall states:
\begin{gather}
\sigma^{\rm cf}_{xy} = \left\{\begin{matrix}
-\frac{1}{4\pi}, & b_0 < 0, \mu> 0 \\ \frac{1}{4\pi}, & b_0 > 0, \mu< 0
\end{matrix}\right.
\end{gather}
Again, we find that IQHIT displays self-duality as one goes from a theory of electrons to that of composite fermions with a change of sign of the Hall conductivity. In certain sense, this self-duality is more robust in Son's theory since it contains an explicit particle-hole symmetry transformation. While, the HLR theory appears to require a long range disorder (Eq. \eqref{compressibility}). 

Lastly, the non-linear sigma model for Dirac composite fermions is identical to that of $g=2$ theory (Eq. \eqref{NLSM}). We have already derived it in appendix \ref{topological_term_derivation}.

\bibliography{bigbib}

\end{document}